\documentclass[english,aps,prper,reprint,showpacs,titlepage,longbibliography]{revtex4-2}   

\usepackage[T1]{fontenc}	% should generally be included for better accented-word behavior
\usepackage[latin9]{inputenc}	% should generally be included for better accent behavior
\usepackage{geometry}		% for controlling page margins
\usepackage{graphicx}
\usepackage[above,below]{placeins}	% allows use of \FloatBarrier command to force section barriers
\usepackage{times}

\usepackage{hyperref}  
\hypersetup{colorlinks=true,urlcolor=blue,citecolor=blue,linkcolor=blue}   
\usepackage{enumitem}          % package for handling list formatting
\setlist{nosep}                 % Tightest spacing for lists. `noitemsep` is more relaxed

\usepackage{amsfonts}

\usepackage{xcolor}
\definecolor{commentblue}{rgb}{0,0,1}

\begin{document}

\begin{titlepage}

  \title{Reflecting to learn in a physics multimedia communication course}

  \author{Steven W. Tarr}
  \affiliation{School of Physics, Georgia Institute of Technology, 837 State Street, Atlanta, Georgia 30332, USA} 
  \author{Emily \surname{Alicea-Mu\~{n}oz}}
  \affiliation{School of Physics, Georgia Institute of Technology, 837 State Street, Atlanta, Georgia 30332, USA}  

  % \keywords{}

  \begin{abstract}
    Science communication skills are considered essential learning objectives for undergraduate physics students. However, high enrollment and limited class resources present significant barriers to providing students ample opportunities to practice their formal presentation skills. We investigate the use of integrated critical reflection and peer evaluation activities in a physics senior seminar course both to improve student learning outcomes and to supplement highly restricted presentation time. Throughout the semester, each student delivers one 8-min multimedia presentation on either their research or an upper-division course topic. Following each presentation, audience members complete one of two randomly assigned peer evaluations: a treatment form that prompts critical reflection or a control form that does not. Each class period concludes with a short quiz on concepts presented in that day's presentations. We observe minimal differences in quiz scores between students in the control and treatment groups. Instead, we find that retention and transfer of presentation content correlate with certain metrics of presentation quality described in the Cognitive Theory of Multimedia Learning and with self-identified prior exposure to presentation topics. \clearpage
  \end{abstract}
  %% Adding the `\clearpage` is the hack to make the title page.  In 2020, the proceedings is
  %% going to be double blind.  This change makes it so that we can programmatically remove the
  %% title page.  In the future, other blinding measures should be taken as well (for example,
  %% removing self-citations).  This is not needed in 2019.

  \maketitle
\end{titlepage}

\section{Introduction}

Science communication skills are considered to be foundational learning objectives for physics students~\cite{NationalCommitteeonScienceEducationStandardsandAssessment1996,Kozminski2014,Yore2004,Hinko2014}. However, the implementation of dedicated science communication courses has been slow and highly localized. Many physics departments instead incorporate occasional structured communication practice within non-communication courses~\cite{Doumont2002,Stewart2010,Holmes2020,Rethman2021}. Others rely on students absorbing these skills through observation and unstructured practice~\cite{Hwang2011,Rodriguez2012,Burkholder2022}. Nevertheless, employer accounts highlight a widespread belief that physics graduates are content specialists with noticeable deficiencies in social and communicative skills~\cite{Chapman1952,HartResearchAssociates2013,Sarkar2016}.

Though many physics educators support direct instruction and student practice of science communication skills, scalability challenges and limited classroom resources often impede deliberate efforts to implement these learning objectives. To better understand how our students develop science communication skills in a restricted academic environment, we conducted a study on a one-credit-hour oral communication Senior Seminar course required for all physics majors at the Georgia Institute of Technology (GT). With high enrollment (typically 40+ per semester) and low contact hours ($<$~1/week), the course lacks sufficient time for each student to present more than once. However, this system provides students with ample opportunities to observe and critically engage with diverse peer presentations to develop their own presentation skills and styles. To assess the value of this course structure, we ask our first research question: \textit{Does reflection following observation of peer oral presentations in a physics classroom affect students' growth as science communicators?}

Understanding the development of science communication skills requires a metric for communication quality. Mayer's Cognitive Theory of Multimedia Learning (CTML) provides a potential framework for understanding the efficacy of varied communication methods~\cite{Mayer2020}. Expanding on Cognitive Load Theory~\cite{Sweller1994}, CTML aims to explain how multimedia affects the interaction between cognitive load and working memory as well as the potential of multimedia lessons to enhance learning gains. Mayer codifies key results from CTML research as fifteen principles of multimedia design~\cite{Mayer2020}, seven of which are accessible in the context of our seminar course (Table~\ref{table:Principles}). The Coherence and Redundancy principles both focus on removing unnecessary information such as intriguing yet irrelevant details and redundant on-screen text respectively. The Signaling and Spatial Contiguity principles emphasize achieving visual clarity by steering learners through otherwise dense presentations and positioning corresponding text and pictures near one another respectively. The Modality principle encourages presenters to explicitly call out and narrate on-screen graphics rather than requiring that observers read bulky captions to understand images. The Personalization and Embodiment principles highlight the importance of cultivating a casual, motivational learning environment through informal, jargon-free language and a spirited, active presence respectively.

\begin{table*}[htbp]
  \caption{Effect size $d$ of adherence to multimedia design principles on quiz performance alongside corresponding median results $d_{\text{lit}}$ from established literature. $n_1$ and $n_0$ represent the number of students who observed and were quizzed on a presentation that did or did not follow a principle respectively. We only report data for the seven principles for which we obtained sufficiently diverse samples (i.e., $n_0, n_1 > 100$). \label{table:Principles}}
  \begin{ruledtabular}
    \begin{tabular}{lcccc}
      \textbf{Multimedia Design Principle} & $n_0$ & $n_1$ & $d$ & $d_{\text{lit}}$~\cite{Mayer2020}\\ 
      \hline
      \textbf{Coherence (Co):} Omit extraneous, seductive details. & $671$ & $362$ & $0.14^{*}\pm 0.07$ & $0.86$\\
      \textbf{Signaling (Sg):} Visually guide learners through content organization. & $782$ & $251$ & $0.13\pm 0.07$ & $0.70$\\
      \textbf{Redundancy (Re):} Avoid text that is redundant with narration or images. & $657$ & $376$ & $-0.25^{***}\pm 0.06$ & $0.72$\\
      \textbf{Spatial Contiguity (SpCt):} Place corresponding slide contents near each other. & $222$ & $811$ & $-0.22^{**}\pm 0.08$ & $0.82$\\
      \textbf{Modality (Md):} Complement graphics with narration, not blocks of text. & $453$ & $580$ & $0.030\pm 0.063$ & $1.00$\\
      \textbf{Personalization (Pn):} Use a conversational, informal style. & $415$ & $618$ & $0.53^{***}\pm 0.06$ & $1.00$\\
      \textbf{Embodiment (Em):} Augment instruction with dynamic, physical expression. & $395$ & $638$ & $0.014\pm 0.064$ & $0.58$\\
    \end{tabular}
  \end{ruledtabular}
  $^{*}p < 0.05;\hspace{1mm}^{**}p < 0.01;\hspace{1mm}^{***}p < 0.001$
\end{table*}

CTML principles have gained traction in Mayer's home field of psychology over the past 30 years~\cite{Noetel2022}. However, the few studies on their applications in physics classrooms have demonstrated inconsistent results, raising doubts on the transferability of these principles~\cite{Chen2012,Chen2014,Wu2015,Magana2019,Morphew2020}. For example, CTML struggles to clearly distinguish multiple harmful and helpful visual representations, and many experiments used to validate its principles employed prerecorded, heavily scripted presentations with 8--10 s per slide in research-laboratory settings inauthentic to the classroom. In pursuit of a research-validated approach to successful science communication, we ask our second research question: \textit{What principles, if any, of CTML apply in authentic physics classroom contexts?} 

\section{Methods}

Before any data collection began, we obtained approval from GT's Institutional Review Board, IRB Protocol H23266. Our study spanned the Fall 2023 (F23, $N_{S1} = 32$) and Spring 2024 (Sp24, $N_{S2} = 17$) offerings of the seminar course, which were taught by different instructors. Both classes met for one 50-min period weekly until each enrolled student completed one presentation. In F23, students self-selected their presentation dates; in Sp24, the instructor randomly assigned dates. In-class survey results indicate that all participants were upper-division physics majors with highly varied classroom science communication experience and minimal professional communication experience.

Excepting the first day of class during which the instructor reviewed the course syllabus and learning outcomes, a typical class period consisted of up to four student presentations. Presentations were capped at 8 min and followed by a 2-min question-and-answer period. Students were instructed to choose a presentation topic befitting an upper-division undergraduate physics audience. Common topics included ongoing student research and past summer internships, though students could instead present a journal article or upper-division course topic. We attended and privately evaluated each presentation based on its adherence or non-adherence to the seven accessible CTML principles. To ensure observation of an authentic representation of the course, neither the students nor the instructors received information on CTML, nor were they made aware of the details of our rubric. Following each presentation, audience members completed non-anonymous peer evaluations counted for an attendance grade. The instructor also provided detailed feedback to each presenter privately after class.

The 2023--24 academic year was the first in 14 years to have two offerings of the senior seminar course. The resultant lower enrollment in Sp24 afforded the instructor flexibility to incorporate brief instruction at the start of the semester. In week 1, the instructor followed the syllabus review with a discussion on science presentation skills based on \textit{Trees, maps, and theorems}~\cite{Doumont2009}. In weeks 2--4, the instructor led in-class workshops to help students design, iterate, and practice their presentations. Students came to class each week having completed a mandatory checkpoint assignment to prepare for the workshop. While students worked in rotating groups of three or four, the instructor moved between groups, asking probing questions, offering personalized feedback, and occasionally addressing the whole class with general comments. Starting in week 5, students began delivering their presentations, and the rest of the Sp24 semester proceeded identically to F23.

Inspired by Girard's work on the perceptions and benefits of peer evaluations following oral presentations~\cite{Girard2011}, we framed our study around the course's peer evaluations. Each student filled one of two peer evaluation forms (internally labeled treatment and control) per presentation. We created a randomized control trial by shuffling the forms prior to distribution. Both forms contained five questions: two 4-pt Likert-type, one check-all-that-apply (CATA), and two short-response. The CATA question asked students where they had previously encountered the presentation topic to establish prior knowledge. The treatment form prompted students to critically reflect by evaluating the presentation content and quality. Students also listed two content items they learned or thought were presented well and two techniques that contributed to the presentation quality. The listing prompt constitutes a form of active learning and has been shown to improve student listening skills~\cite{Girard2011}. The control form eschewed critical reflection, instead prompting students to evaluate the presentation's audibility and legibility. Students also briefly described their past experiences and engagement with the presentation topic and how it compared to their current engagement. Though the control form masqueraded under the pretense of evaluating the social components of a presentation, it was designed to keep students from engaging in any meaningful reflection on the presentation content.

Following the final peer evaluation per class period, we administered a short quiz graded on participation to all students. Two days before their own presentation, each student submitted a few questions on their presentation topic that they expected an attentive audience member could answer. The instructor reinforced throughout the course that students should write the questions first, design their presentation around the questions, and revise both as needed. We asked students to categorize their questions as pertaining either to material explicitly presented (i.e., retention) or to the synthesis of presented materials into a new idea (i.e., transfer). The instructor vetted each question and provided feedback if necessary. We built the quiz by randomly selecting one question from each presenter such that there were no more than two retention questions and two transfer questions per quiz. Independent of course grades, we scored each question on the following rubric: no attempt/fully incorrect (0 pt), partially correct (0.5 pt), and fully correct (1 pt). We hypothesized that students in the treatment group would score higher on average because of the critical reflection prompt~\cite{Girard2011,Handayani2019}.

\section{Results}

% \begin{figure}[b]
%   \includegraphics[width=0.65\linewidth]{Fig2_CTMLDistribution.png}
%   \caption{Presentation quality roughly follows a normal distribution in both semesters. Though the semester means differed negligibly, the reduced variance in the spring semester may be owed to the addition of brief instruction on presentation standards.
%   \label{fig:Histogram}}
% \end{figure}

\subsection{Students as Presenters}

\begin{figure}[b]
  \includegraphics[width=\linewidth]{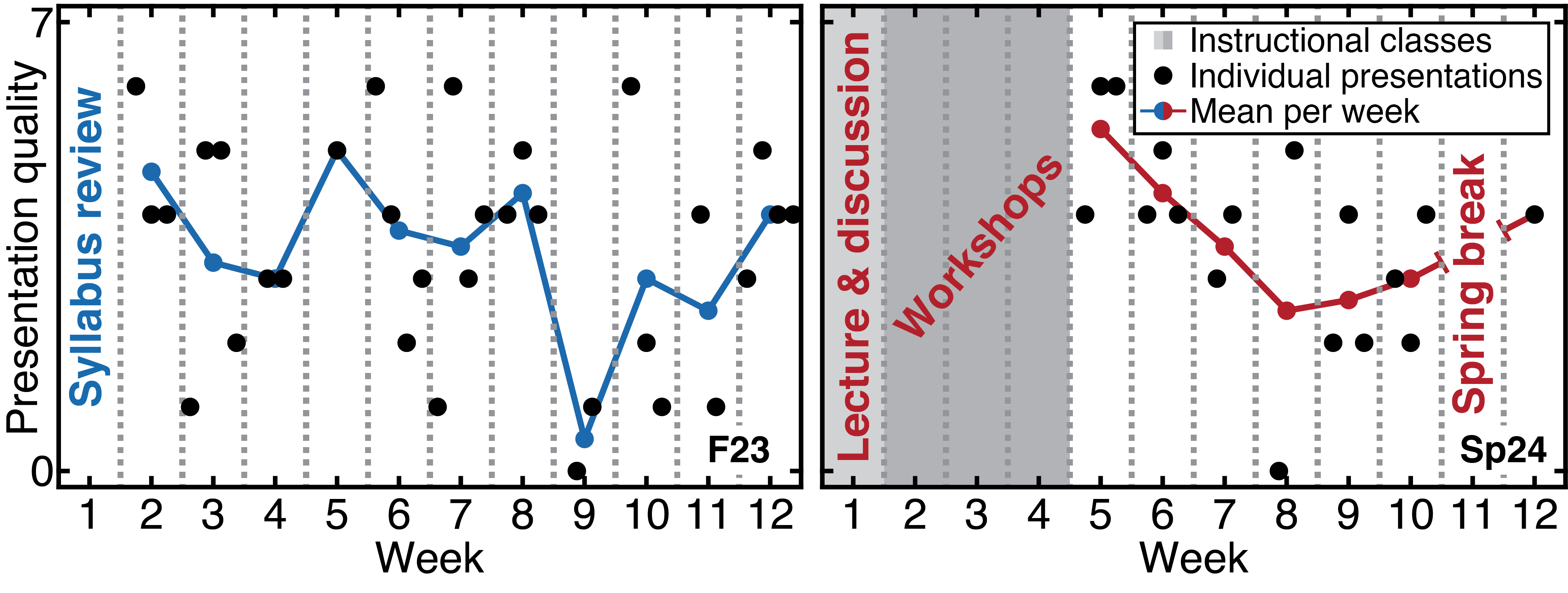}
  \caption{Presentation quality as measured by adherence to CTML principles remains roughly static throughout both semesters.\label{fig:CTMLvsTime}}
\end{figure}

We report our analysis of the seven accessible CTML principles in Table~\ref{table:Principles}. To investigate trends in presentation quality, we used a proxy variable given by the total number of CTML principles each student obeyed. We observed presentation quality to stay roughly constant with a slight decline throughout either semester (Fig.~\ref{fig:CTMLvsTime}). Distributions of presentation quality were centered about the midpoint of the rubric ($M_{S1}=3.5,~SD_{S1}=1.7;~M_{S2}=3.6,~SD_{S2}=1.5$). A one-sided Mann-Whitney \textit{U} test does not suggest a significant improvement due to four weeks of instruction in Sp24 ($z = -0.30,~p = 0.38$). The Cohen's $d$ statistic reinforces this notion; we found the instruction to have an inconclusive effect on presentation quality ($d = 0.11\pm 0.30$).

Because the seminar course lacks the time for multiple student presentations, the observed quality-time correlation may not properly represent student growth through observation or reflection. We can refine this result by substituting the number of control or treatment forms a student fills before finalizing their presentation in place of time. These new proxy variables better represent how students engage with peer presentations and account for when students present relative to each other. Still, we found a similar decline in presentation quality across both semesters as students completed more of either peer evaluation form. Spearman's $\rho$ indicates weak negative correlations between presentation quality and completion of control ($\rho = -0.10 \pm 0.18,~p = 0.60$) and treatment ($\rho = -0.26 \pm 0.18,~p = 0.15$) forms in F23. We observed stronger negative correlations in Sp24 (control: $\rho = -0.51 \pm 0.22,~p = 0.04$; treatment: $\rho = -0.52 \pm 0.22,~p = 0.03$), possibly owed to smaller sample size.

\begin{figure}[t]
  \includegraphics[width=\linewidth]{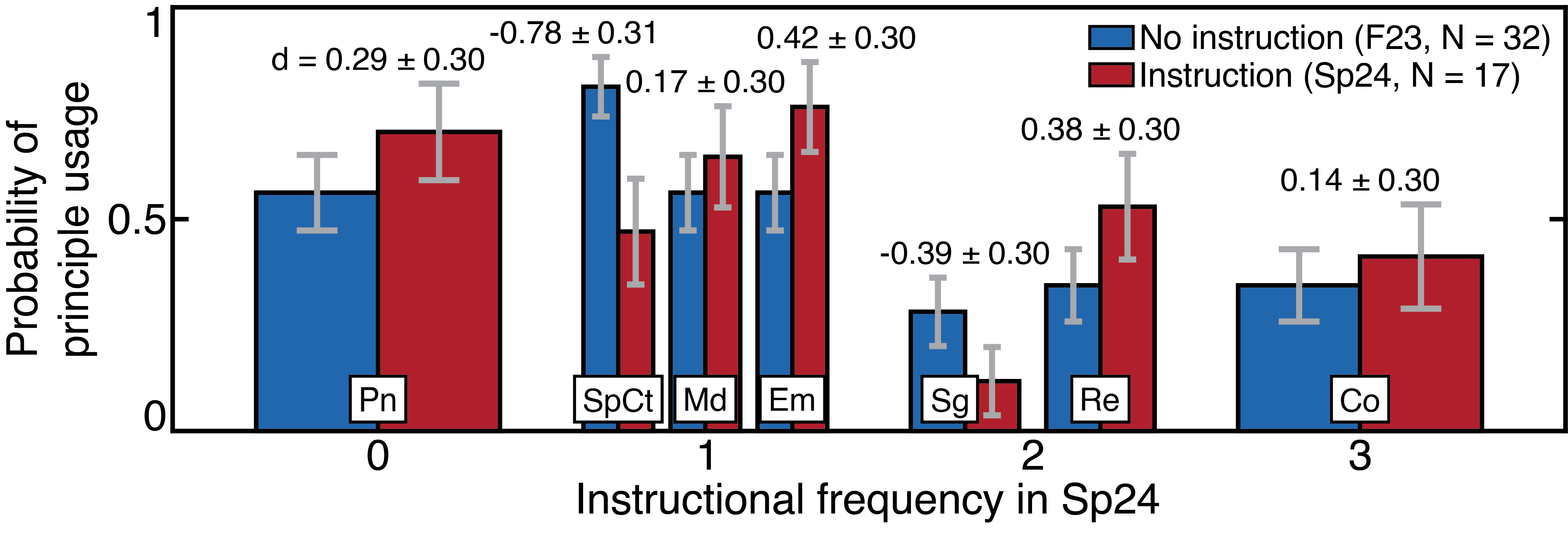}
  \caption{Although students in Sp24 received instruction corresponding to specific CTML principles, their use of those principles was similar to that of students who received no instruction.
  \label{fig:Instruction}}
\end{figure}

Though the instructors were not explicitly informed on CTML, their experiences as researchers, professors, and science communicators caused some overlap between their feedback to students and CTML principles. To further probe the effect of early instruction on presentation design and delivery, we mapped the Sp24 instructor's lecture and workshop notes onto corresponding CTML principles. For each semester, we plotted a class-wide probability of principle usage with principles ordered by instructional frequency (Fig.~\ref{fig:Instruction}). Though the Sp24 data suggests that principle adherence decreases with increasing instructional frequency, comparison with F23 students who received no instruction reveals the perceived effect to be coincidental. The effect size of instruction varied greatly between principles regardless of observed instructional frequency and often fell within one standard error of zero. We note the moderate difference in usage of the Personalization principle, which was not taught in either semester; this suggests a possible underlying difference between the F23 and Sp24 students that may partially account for the disparity between semester means. Nevertheless, a one-sided Mann-Whitney \textit{U} test did not suggest improvement in usage of any observed CTML principles due to instruction alone.

\subsection{Students as Observers}

\begin{figure}[t]
  \includegraphics[width=\linewidth]{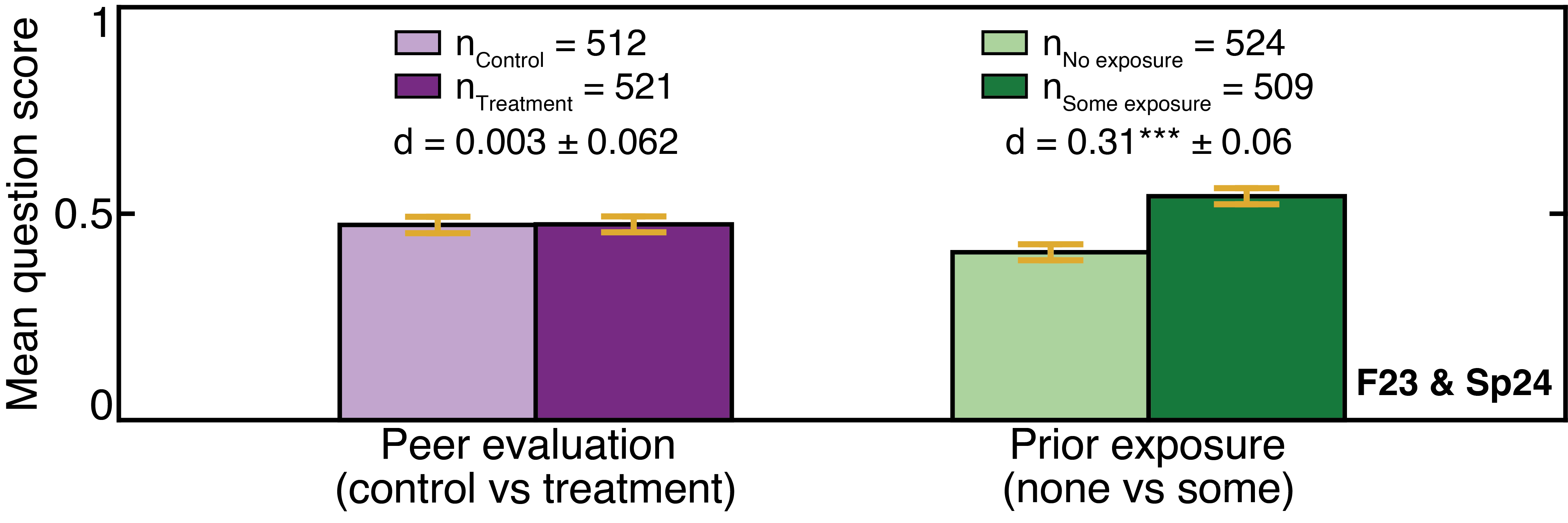}
  \caption{We observed negligible differences in quiz performance between students who did or did not engage in reflection directly following each presentation. Instead, students who self-identified as having prior exposure to presentation concepts significantly outperformed those with no prior exposure.
  \label{fig:Quiz}}
\end{figure}

Having established presenter trends, we now turn students' roles as audience members observing and reflecting upon peer presentations. Barring absences, each student observed each of their classmates' presentations and answered one question per presentation; consequently, we graded 1033 responses across 47 questions and compared mean question scores across two categories (Fig.~\ref{fig:Quiz}). We did not observe a statistically significant effect due to the critical reflection activity (Cohen's $d = 0.003 \pm 0.062,~p = 0.47$). However, students who indicated having some prior exposure to presentation content (e.g., college course, internship, self study) moderately outperformed students who indicated none, with high statistical significance ($d = 0.31\pm 0.06,~p < 0.001$).

By separating question responses based on the corresponding presentation's adherence to specific CTML principles, we measured the effects of each individual principle on students' retention and transfer of presentation concepts. We report our statistics for each principle including sample sizes, mean scores, and effect size alongside Mayer's established effect sizes from Ref.~\cite{Mayer2020} in Table~\ref{table:Principles}. Across all seven principles, we observed noticeably smaller effect size magnitudes than Mayer. Our students benefited most from observing presentations that followed the Personalization principle. We also observed the Coherence and Signaling principles to have small, positive effects on quiz scores. Unlike in Mayer's work, we found negligible effect on retention or transfer due to Modality or Embodiment. Further, we observed moderately small negative effects from adherence to the Redundancy and Spatial Contiguity principles.

\section{Discussion}

Barriers to student development into skilled science communicators are readily apparent in this course. Without the opportunity to deliver multiple authentic presentations and put instructor and peer feedback into practice, students are greatly limited in the scope of their growth. Ideally, the audience-centered circumstances of this course would encourage the development of science communication skills through repeated critical reflection on the many peer presentations each student observes, but our data suggests otherwise. Quiz scores were nearly indistinguishable between groups of students who engaged in reflection activities and groups who did not. Presentation quality as measured by total CTML principles followed did not improve with increased peer evaluation over either semester. \textit{Student presentations appear to be unaffected by prior observation of and reflection on peer presentations in the context of this course.}

Despite the flexibility to include instruction in Sp24, we observed only incremental improvements in student learning outcomes. Students in Sp24 delivered presentations that followed CTML principles nearly identically to those in F23 regardless of instructional frequency. In one-on-one, semi-structured interviews conducted after the conclusion of the Sp24 course, students from the F23 class regularly lamented the lack of instructional resources. Though students from the Sp24 class expressed appreciation for the brief instruction they received, they often voiced yearning for more involved instruction throughout a longer portion of the semester.

We observed substantial differences between effect sizes of CTML principles tested in this course and those reported by Mayer. For five of the seven tested principles, effects remained positive but were significantly reduced. We speculate that these smaller effects may be due to our use of an authentic classroom environment compared to Mayer's primarily research-laboratory-based environments. Students in the classroom often have multiple priorities competing for their attention that may not be present in the research laboratory, such as intrinsic and extrinsic pressures to attain high grades. These additional priorities may further limit the cognitive capacity that students can dedicate to retention and transfer of peer presentation content, resulting in muted effect sizes for each principle.

For two of the seven principles, we observed negative effects on quiz scores. For the Redundancy principle, this result implies that students performed better on questions corresponding to presentations where the speaker included on-screen text redundant to graphics or narration. This effect might arise from the course's substantial intrinsic cognitive load. Our students rapidly experienced a wide variety of often unrelated and unfamiliar upper-division and research-level topics. We surmise that one such consequence is a reversal of the Redundancy principle, where reinforcement through redundant on-screen text can become helpful for retention or transfer. Indeed, perspectives in the PER community often emphasize the importance of multiple overlapping visual representations to improve student outcomes~\cite{Chen2014,Opfermann2017}.

A negative effect size for the Spatial Contiguity principle implies that students performed better on questions for presentations where corresponding on-screen text and pictures were not placed close together. This effect may be the result of students observing the presentations on the large screen of a lecture hall. At such scales, distances between graphical elements may play different roles in student cognition than on typical computer screens. Another potentially confounding effect may be the principle's definitional ambiguity. What does it mean for corresponding materials on a slide to be ``presented near rather than far from each other''~\cite{Mayer2020}? We expect our forthcoming interrater agreement statistics to be helpful in identifying any especially problematic principle definitions. \textit{Ultimately, the stark differences between our effect sizes and Mayer's further the developing narrative that there may be boundary conditions for CTML principles specific to the physics classroom}~\cite{Chen2012,Chen2014,Wu2015,Opfermann2017}. 

Our study possesses notable limitations resulting from the nature of observational student data and small sample size. Consequently, it becomes difficult to examine all potentially relevant confounding variables or to generalize our conclusions. The above-mentioned barriers to supporting course learning objectives also apply as constraints on testing substantive changes to the course structure. In turn, we began a study in Sp24 of an analogous two-credit hour chemistry Senior Seminar course that offers more rigorous instruction and multiple chances to present. Within both the physics and chemistry courses, we plan to perform qualitative analyses on written reflections, instructor feedback, and student interviews to develop a more comprehensive understanding of how students interact with dedicated science communication courses.

Within our analysis, we observed limitations of using the number of CTML principles obeyed as a proxy variable for presentation quality. One plausible element of presentation quality is how well the audience learns from the presentation. If quiz performance accurately reflects the learning that results from each presentation, then the total principles followed is a poor proxy for presentation quality, as it only weakly correlates with mean question score (Spearman's $\rho = 0.17 \pm 0.14,~p=0.24$). In turn, we have begun exploring other potential proxy variables that necessarily correlate with audience understanding. Finally, in our goal to understand the applications of CTML in physics, we have begun to encounter limitations in its transferability. We are working to expand our investigation to include assessments of alternative interpretations of multimedia learning such as Schnotz's Integrative Model of Text-Picture Comprehension~\cite{Schnotz2022}.

\acknowledgments{We thank physics instructors Colin Parker and Itamar Kimchi and chemistry instructor Mary Peek for their support throughout our research studies on their courses. We thank Michael Schatz, Edwin Greco, and the rest of the PER group at GT for helpful discussions. This work is partly supported by School of Physics seed funds and the GT College of Sciences Teaching Effectiveness, Advocacy and Mentoring (TEAM) Committee funds.}

\end{document}